\documentclass[prd,showpacs,preprintnumbers,groupedaddress]{revtex4-1} 

\usepackage{amsthm}
\theoremstyle{plain}
\newtheorem{theorem}{Theorem}
\newtheorem{definition}{Definition}

\newtheorem{corollary}{Corollary}
\newtheorem{proposition}{Proposition}
\usepackage{cases}
\usepackage{color}
\usepackage{comment}
\usepackage[normalem]{ulem}
\usepackage{amsfonts}

\newcommand{\llaabel}[1]{\label{#1}}

\newcommand{\red}[1]{\textcolor{red}{#1}}

\begin{document}


\title{Stability of null orbits on photon spheres and photon surfaces}


\author{Yasutaka Koga}
\author{Tomohiro Harada}
\affiliation{Department of Physics, Rikkyo University, Toshima, Tokyo 171-8501, Japan}


\date{\today}

\begin{abstract}
Stability of a photon sphere, or stability of circular null geodesics on the sphere, plays a key role in its applications to astrophysics.
For instance, an unstable photon sphere is responsible for determining the size of a black hole shadow, while a stable photon sphere is inferred to cause the instability of spacetime due to the trapping of gravitational waves on the radius.
A photon surface is a geometrical structure first introduced by Claudel, Virbhadra and Ellis as the generalization of a photon sphere.
The surface does not require any symmetry of spacetime and has its second fundamental form of pure trace.
In this paper, we define the stability of null geodesics on a photon surface.
It represents whether null geodesics perturbed from the photon surface are attracted to or repelled from the photon surface.
Then, we define a strictly (un)stable photon surface as 
a photon surface on which all null geodesics are (un)stable. 
We find that the stability is determined by Riemann curvature.
Furthermore, it is characterized by the normal derivative of the second fundamental form.
As a consequence, for example, a strictly unstable photon surface requires nonvanishing 
Weyl curvature on it if the null energy condition is satisfied.
\end{abstract}


\maketitle


\tableofcontents

\section{Introduction}
\llaabel{sec:introduction}
A photon sphere is a sphere of spacetime on which null geodesics take circular orbits.
In astrophysical cases, black holes usually have photon spheres near their horizons.
A photon sphere has been widely studied in its various aspects;
for optical observations of black holes through background light emission, the photon sphere is related to the size of the black hole shadow.
In the case of the Schwarzschild black hole, for example, we can see their relation from the calculation  by Synge~\cite{synge}.
Quite recently, the Event Horizon Telescope Collaboration has observed, for the first time, the shadow of the supermassive black hole candidate in the center of the galaxy M87 and derived its mass by comparing the images with the theoretical expectations~\cite{eht}; 
properties of gravitational waves from black holes are also closely related to the photon sphere.
It is known that the frequencies of quasinormal modes are related to the parameters of null geodesic motions on and near the photon sphere in
various situations~\cite{cardoso}~\cite{hod}.
\par
Stability of a photon sphere, i.e. stability of the circular orbits on the sphere, plays key roles in the applications of a photon sphere to astrophysics.
For instance, the photon sphere that shapes the black hole shadow is unstable.
A stable photon sphere, on the other hand, is inferred to cause instability of spacetime~\cite{keir_2014}~\cite{cardoso_2014}~\cite{cunha}.
When spacetime is perturbed, gravitational waves propagating nearly along a stable photon sphere would grow nonlinearly, while they are trapped and coupled with each other in the vicinity of the radius.
They will probably break the structure near the sphere and, finally, the spacetime.
In fluid dynamics on curved spacetime, the stability of a photon sphere also has remarkable importance.
Recently, it has been found that radiation fluid flow has its sonic point only on an unstable photon sphere~\cite{koga}~\cite{koga2}~\cite{koga3}.
This surprising phenomenon, named sonic point/photon sphere correspondence, appears in quite various situations and provides examples where a photon sphere plays an important role in non-null motion of matter.
\par
A photon surface is a geometrical structure first introduced by Claudel, Virbhadra and Ellis~\cite{claudel} as the generalization of a photon sphere.
The surface is defined so that it inherits only the local properties of a photon sphere and does not necessarily have symmetries.
Together with the definition, the authors also proved a theorem concerning the equivalent conditions for a surface to be a photon surface as one of the main results.
The theorem (Theorem~2.2 in~\cite{claudel}) states that a given timelike hypersurface is a photon surface if and only if it is totally umbilic, i.e. the second fundamental form is pure trace everywhere.
Subsequently, Perlick~\cite{perlick} proved that the theorem holds for arbitrary dimensions of the surface and the spacetime.
Since a photon surface requires no symmetries, it would have applicability to many physical problems in addition to its own interest as a geometrical object.
\par
As with the stability of a photon sphere, the stability of a photon surface should be also important for the applications of a photon surface to various problems of physics.
In this paper, we define the stability of null geodesics along a photon surface and derive the stability conditions.
Usually, the stability of a photon sphere is easily defined because the null geodesic in static and spherically symmetric spacetime obeys a one-dimensional equation of motion and the problem reduces to analyzing the effective potential.
In particular cases, the stability of photon surface was defined by use of optical metric~\cite{gibbons_2016} and by the effective potential method~\cite{koga3}.
For a generic photon surface, we define the stability in a covariant manner by considering a geodesic deviation.
Since a geodesic deviation is governed by a local geometrical quantity, Riemann curvature, the stability condition is finally obtained in terms of the curvature.
\par
This paper is organized as follows.
In Sec.~\ref{sec:stability-ps}, we review and reinterpret the stability of a photon sphere.
We see how the stability is expressed by a geodesic deviation.
In Sec.~\ref{sec:stability-psf}, we define the stability of null geodesics along a photon surface based on the arguments for a photon sphere and derive the stability condition in terms of Riemann curvature.
In sec.~\ref{sec:another}, we derive an alternative expression of the stability condition in terms of the second fundamental form with an appropriate foliation, and give another interpretation of our definition of stability.
The stability conditions in Secs.~\ref{sec:stability-psf} and~\ref{sec:another} are guaranteed to be equivalent by Raychaudhuri equation for the unit normal vector field of the foliation.
The stability conditions indicate that we can a priori identify the stability before finding photon surfaces of spacetime explicitly.
For example, any photon surface in conformally flat spacetime is stable if the null energy condition is satisfied.
We see the corollaries for such special cases in Sec.~\ref{sec:corollaries}.
The conclusion is given in Sec.~\ref{sec:conclusion}.

\section{Stability of Photon Sphere}
\llaabel{sec:stability-ps}
Consider static and spherically symmetric spacetime.
A hypersurface of constant radius, $S=\mathbb{R}\times S^2$, is called a photon sphere if there exist null circular orbits, i.e. null geodesics whose spatial orbits are circles, on $S$.
The photon sphere is said to be stable if the circular orbits are stable circular orbits and unstable if unstable circular orbits.
We can describe the stability in a covariant manner as follows.
\par
For a stable photon sphere, if a null geodesic on the sphere is perturbed from the sphere, the perturbed geodesic is attracted to (accelerated toward) the unperturbed geodesic.
On the other hand, the perturbed geodesic is repelled from (accelerated fromward) the unperturbed geodesic if the photon sphere is unstable.
Therefore, the stability of a null circular geodesic is given by the relative acceleration between the circular geodesic and its infinitesimally nearby null geodesic.
\par
The above argument is represented in terms of a geodesic deviation.
Consider a null circular geodesic $\gamma$ with its tangent vector $k$ on a photon sphere $S$ and the infinitesimally nearby null geodesic $\tilde{\gamma}$ which is obtained by perturbing $\gamma$ in the radial direction at a point $p\in S$.
Let $X$ be the deviation vector arising from $\gamma$ and $\tilde{\gamma}$.
It satisfies the condition $X\propto n$ at $p$ for
the unit normal vector $n$ of $S$.
Then the relative acceleration between $\gamma$ and $\tilde{\gamma}$ is given by $a=\nabla_k\left(\nabla_kX\right)$ and $\gamma$ is stable if $\left.g(X,a)\right|_p<0$ while unstable if $\left.g(X,a)\right|_p>0$.
If $\left.g(X,a)\right|_p=0$, $\gamma$ is marginally stable.
\par
Note that because of the symmetry, if there is a null geodesic $\gamma$ on a photon sphere $S$ that is stable, unstable, and marginally stable at $p$, $\gamma$ is stable, unstable, and marginally stable, respectively, everywhere on $S$ and all other null geodesics on $S$ has the same stability as $\gamma$.
Therefore photon spheres are completely classified into stable, unstable, and marginally stable ones.

\section{Stability of null orbits along Photon Surface}
\llaabel{sec:stability-psf}
Here, after reviewing a photon surface, we define the stability of a photon surface based on the discussion in Sec.~\ref{sec:stability-ps}.
Then we derive the stability condition in terms of curvature.
\subsection{Photon surface}
A photon surface, defined by Claudel et al.~\cite{claudel}, is a hypersurface on which every null geodesic initially tangent to it remains tangent.
This is the generalization of a photon sphere and can be defined for any spacetime, $(M, {\bf g})$, even if the spacetime has no symmetries like spherical symmetry:
\begin{definition}[Photon surface]
\label{definition:photonsurface}
A photon surface of $(M, {\bf g})$ is an immersed, nowhere-spacelike
hypersurface $S$ of $(M, {\bf g})$ such that, for every point $p\in S$ and every null vector ${\bf k}\in T_pS$, there exists a null geodesic $\gamma : (-\epsilon,\epsilon) \to M$ of $(M, {\bf g})$ such that $\dot{\gamma}(0) ={\bf k}, |\gamma|\subset S$.
\end{definition}
The works by Claudel et al.~\cite{claudel} and Perlick~\cite{perlick} give the equivalent condition for a timelike hypersurface to be a photon surface:
\begin{theorem}[Claudel et al. (2001), Perlick (2005)]
\label{theorem:D-dimPhotonsurface}
Let $S$ be a timelike hypersurface of spacetime $(M,g)$ with $\dim M\ge3$.
Let $n$, $\chi_{ab}$, $\Theta$ and $\sigma_{ab}$ be the unit normal, the second fundamental form, the trace and the trace-free part of $\chi_{ab}$, respectively.
Then $S$ is a photon surface if and only if it is totally umbilic, i.e.
\begin{equation}
\llaabel{eq:umbilic}
\sigma_{ab}=0\;\;\; \forall p\in S.
\end{equation}
\end{theorem}
Note that any null hypersurface is trivially a photon surface~\cite{claudel}.
\subsection{Stability of null geodesics on a photon surface}
Following the argument in Sec.~\ref{sec:stability-ps}, we define the stability of a null geodesic $\gamma$ on a photon surface $S$ in terms of the deviation vector orthogonal to $S$.
The deviation is interpreted as what gives the perturbation of $\gamma$ from $S$:
\begin{definition}
\llaabel{definition:stability-gamma}
Let $S$ be a timelike photon surface of $(M,g)$ and $n$ be the unit normal vector of $S$.
Let $\gamma$ be a null geodesic on $S$ passing a point $p\in S$ and $k$ be the tangent vector to $\gamma$.
Let $X_{k_p}$ be the deviation vector of $\gamma$ satisfying the condition,
\begin{equation}
\left.X_{k_p}\right|_p\propto \left.n\right|_p.
\end{equation}
The null geodesic $\gamma$ is said to be stable, unstable, and marginally stable at $p$ if the acceleration scalar $a_{k_p}:=g\left(X_{k_p},\nabla_k\left(\nabla_kX_{k_p}\right)\right)$ satisfies
\begin{equation}
\llaabel{eq:stability-orbit}
\left.a_{k_p}\right|_p<0,\ \ >0,\ \ \ and\ =0,
\end{equation}
respectively.
\end{definition}
The spacetime dimension is implicitly assumed to be $\dim M\ge3$ since a photon surface in spacetime with $\dim M=2$ is one-dimensional, i.e. a null geodesic itself, and cannot be timelike.
The deviation vector $X_{k_p}$ is, usually, physically interpreted as what gives the null geodesic $\tilde{\gamma}$ which is obtained when $\gamma$ is perturbed at $p$ in the direction orthogonal to $S$.
$a_{k_p}$ represents the relative acceleration of $\tilde{\gamma}$ to $\gamma$, or $S$.
This is what we need for our description of the stable or unstable behaviors of (perturbed) null geodesics, being attracted to or repelled from $S$.
Note that $\gamma$ can be either stable or unstable depending on the point $p$.
Furthermore, the stability also depends on the direction $k\in T_pS$ of the null geodesic.
If $\gamma$ is stable, unstable, and marginally stable at $\forall q\in|\gamma|$, we simply call it stable, unstable, and marginally stable, respectively.
\par
From the geodesic deviation equation, the left-hand side of Eq.~(\ref{eq:stability-orbit}) is calculated as
\begin{eqnarray}
\left.a_{k_p}\right|_p
&=&-\left[R_{acbd}X_{k_p}^ak^cX_{k_p}^bk^d\right]_p\nonumber\\
&=&-\left[X_{k_p}^2R_{acbd}n^ak^cn^bk^d\right]_p
\end{eqnarray}
where $X_{k_p}^2:=g_{ab}X_{k_p}^aX_{k_p}^b$ is positive.
Then we reach the following stability condition:
\begin{proposition}
\llaabel{proposition:stabilitycondition-curvature}
\llaabel{proposition:stability-curvature}
Let $S$ be a timelike photon surface and $\gamma$ be a null geodesic on $S$ with the tangent vector $k$ at $p\in S$.
Then $\gamma$ is stable, unstable, and marginally stable at $p$ if and only if
\begin{equation}
\llaabel{eq:stabilitycondition-curvature}
R_{acbd}k^an^ck^bn^d>0,\;<0,\; and\; =0,
\end{equation}
respectively, at $p$.
\end{proposition}
It is worth noting that the component $R_{acbd}k^an^ck^bn^d$, or more generally $R_{ecfd}h^e_an^ch^f_bn^d$ where $h_b^a$ is the induced metric on $S$, is the missing component in Gauss-Codazzi equations for the decomposition of the curvature concerning $S$ and $n$.
Therefore it cannot be expressed solely in terms of the intrinsic and extrinsic curvatures~\cite{textbook:poisson}.
\par
The decomposition of Riemann tensor into Weyl tensor $C_{abcd}$ and Ricci tensor $R_{ab}$ often helps us to understand the physics.
We also have the expression alternative to Proposition~\ref{proposition:stability-curvature}:
\begin{proposition}
\llaabel{proposition:stability-weyl}
Let $S$ be a timelike photon surface and $\gamma$ be a null geodesic on $S$ with the tangent vector $k$ at $p\in S$.
Then $\gamma$ is stable, unstable, and marginally stable at $p$ if and only if
\begin{equation}
\llaabel{eq:stabilitycondition-weyl}
C_{acbd}k^an^ck^bn^d+\frac{1}{D-2}R_{ab}k^ak^b>0,\;<0,\; and\; =0,
\end{equation}
respectively, at $p$ where $D\ge3$ is the spacetime dimension.
\end{proposition}
Although a photon surface $S$ of spacetime $(M,g)$ is invariant submanifold under a conformal transformation $(M,g)\to (M,\Omega^2g)$~\cite{claudel}, Proposition~\ref{proposition:stability-weyl} tells us that the stability of $S$ is not conformally invariant due to the presence of Ricci tensor in the stability condition.

\subsection{Stability of a photon surface}
There can be both stable and unstable null geodesics on a photon surface $S$.
We define {\it the stability of a photon surface} in cases where all the null geodesics on $S$ are (un)stable:
\begin{definition}
\llaabel{definition:stability-psf}
Let $S$ be a timelike photon surface and $n$ be the unit normal vector of $S$.
Let $k_p\in T_pS$ be a null vector on a point $p\in S$.
Let $\gamma$ be the null geodesic on $S$ passing $p$ with the tangent vector $k_p$.
The photon surface $S$ is said to be
\begin{itemize}
\item stable if $\left.a_{k_p}\right|_p\le0$ $\forall k_p\in T_pS$, $\forall p\in S$,
\item strictly stable if $\left.a_{k_p}\right|_p<0$ $\forall k_p\in T_pS$, $\forall p\in S$,
\item unstable if $\left.a_{k_p}\right|_p\ge0$ $\forall k_p\in T_pS$, $\forall p\in S$,
\item strictly unstable if $\left.a_{k_p}\right|_p>0$ $\forall k_p\in T_pS$, $\forall p\in S$, and
\item marginally stable if $\left.a_{k_p}\right|_p=0$ $\forall k_p\in T_pS$, $\forall p\in S$,
\end{itemize}
where $a_{k_p}$ is the acceleration scalar defined for $\gamma$ at $p$ as in Definition~\ref{definition:stability-gamma}.
\end{definition}
The left-hand side of the conditions can be expressed in terms of curvatures from Propositions~\ref{proposition:stability-curvature} and~\ref{proposition:stability-weyl}.

\section{Stability and second fundamental form}
\llaabel{sec:another}
With a spacetime foliation, the curvature of spacetime is related to the first derivative of second fundamental forms, which is the tensor constructed from the second derivative of the unit normal vector field.
Therefore, the stability condition in Proposition~\ref{proposition:stability-curvature} can be rewritten in terms of the second fundamental form instead of the curvature.
We here derive the stability condition in terms of the second fundamental form and give another interpretation of the stability, defined in Definition~\ref{definition:stability-gamma}, with a particular spacetime foliation.
\subsection{Stability condition in Gaussian normal foliation}
\llaabel{sec:foliation}
Consider a timelike photon surface $S$ of $(M,g)$ and a spacetime foliation $\left\{S_r\right\}$ in the vicinity of $S$ which includes $S$ as
\begin{equation}
\llaabel{eq:include-psf}
S_{0}=S
\end{equation}
for the parameter $r=0$.
For any foliation, the unit normal vector field $n^a$ of the surfaces generates curves and the congruence consisting of them.
Then the trace-free part of the second fundamental form, $\sigma_{ab}$, of each surface coincides with the shear of the congruence, while the vorticity $\omega_{ab}=0$ by construction.
We identify the shear of the congruence with $\sigma_{ab}$.
Raychaudhuri equation for the congruence gives the relation between the shear evolution and the curvature,
\begin{eqnarray}
\nabla_n\sigma_{ab}
+\sigma_{ac}{\sigma^c}_b
+2\frac{\Theta}{D-1}\sigma_{ab}
+\dot{n}_a\dot{n}_b+\nabla_{(a}\dot{n}_{b)}
-\frac{1}{D-1}h_{ab}\left[\sigma^{cd}\sigma_{cd}+\dot{n}_c\dot{n}^c+\nabla_c\dot{n}^c\right]\nonumber\\
=-R_{acbd}n^cn^d+\frac{1}{D-1}R_{cd}n^cn^dh_{ab},
\end{eqnarray}
where $\dot{n}^a:=\nabla_nn^a$.
On the photon surface $S_{r_p}=S$, the equation reduces to
\begin{equation}
\nabla_n\sigma_{ab}
+\dot{n}_a\dot{n}_b+\nabla_{(a}\dot{n}_{b)}
-\frac{1}{D-1}h_{ab}\left[\dot{n}_c\dot{n}^c+\nabla_c\dot{n}^c\right]
=-R_{acbd}n^cn^d+\frac{1}{D-1}R_{cd}n^cn^dh_{ab}
\end{equation}
from the fact $\sigma_{ab}=0$ $\forall p\in S$.
The left-hand side of the stability condition in Proposition~\ref{proposition:stability-curvature} is therefore rewritten as
\begin{equation}
\llaabel{eq:stability-lhs-shear-generic}
R_{acbd}k^an^ck^bn^d=
-k^ak^b\nabla_n\sigma_{ab}
-k^ak^b\dot{n}_a\dot{n}_b
-k^ak^b\nabla_{a}\dot{n}_{b}
\end{equation}
for any foliation satisfying Eq.~(\ref{eq:include-psf}).
Thus a null geodesic on a photon surface in the direction $k$ at $p$ is stable, unstable, and marginally stable if and only if the right-hand side of Eq.~(\ref{eq:stability-lhs-shear-generic}) is negative, positive, and zero, respectively.
\par
Suppose the foliation in the vicinity of $S$ satisfies the condition,
\begin{equation}
\llaabel{eq:gaussian-foliation}
dn=0,
\end{equation}
for the unit normal $n$ in addition to Eq.~(\ref{eq:include-psf}).
The condition implies
\begin{equation}
\llaabel{eq:geodesic-normal-vector}
n^b\nabla_bn^a=0
\end{equation}
and therefore the parameter $r$ is the one of Gaussian normal coordinates which parametrizes each hypersurface $S_r$.
We refer to the foliation $\left\{S_r\right\}$ satisfying the conditions Eqs.~(\ref{eq:include-psf}) and~(\ref{eq:gaussian-foliation}) as {\it Gaussian normal foliation}.
(One can rescale $r\to r'=r'(r)$ so that $n=dr$, however, here we only assume that the unit normal $n$ points in the same direction as the normal $dr$.)
From Eq.~(\ref{eq:gaussian-foliation}), Eq.~(\ref{eq:stability-lhs-shear-generic}) reduces to
\begin{equation}
R_{acbd}k^an^ck^bn^d=
-k^ak^b\nabla_n\sigma_{ab}
\end{equation}
for the Gaussian normal foliation.
Then we obtain the alternative expression of stability condition in Proposition~\ref{proposition:stability-curvature} in terms of the second fundamental form:
\begin{proposition}
\llaabel{proposition:stability-shear}
Let $S$ be a timelike photon surface and $\gamma$ be a null geodesic on $S$ with the tangent vector $k\in T_pS$ at $p\in S$.
Let $\left\{S_r\right\}$, $\chi_{ab}$, and $\sigma_{ab}$ be Gaussian normal foliation, defined by the conditions in Eqs.~(\ref{eq:include-psf}) and~(\ref{eq:gaussian-foliation}), second fundamental form of each $S_r$, and its trace-free part, respectively.
Then $\gamma$ is stable, unstable, and marginally stable at $p$ if and only if
\begin{equation}
\llaabel{eq:stabilitycondition-shear}
k^ak^b
\left.\nabla_n\sigma_{ab}
\right|_p<0,\;>0,\; and\; =0,
\end{equation}
respectively.
\end{proposition}
A timelike photon surface is a hypersurface characterized by the vanishing of $\sigma_{ab}$.
Similarly, stability of null geodesics on a photon surface is determined by $\nabla_n\sigma_{ab}$.
To identify the stability of a photon surface, it would be easier to calculate the left-hand side of Eq.~(\ref{eq:stabilitycondition-shear}) in Proposition~\ref{proposition:stability-shear} rather than the curvature, Eq.~(\ref{eq:stabilitycondition-curvature}), in Proposition~\ref{proposition:stability-shear} in many cases.
\par
We give the interpretation of Proposition~\ref{proposition:stability-shear} by considering {\it acceleration of a geodesic with respect to a surface} in the following.
\subsection{Acceleration with respect to a hypersurface}
Consider a non-null hypersurface $\mathcal{S}$ of spacetime $(M,g)$ and a (null or non-null) geodesic $\gamma$ which is tangent to $\mathcal{S}$ at a point $p\in \mathcal{S}$ with the tangent vector $v\in T_p\mathcal{S}$.
The tangent vector $v$ to $\gamma$ at $p$ also generates the geodesic $\tilde{\gamma}$ of the subspace $(\mathcal{S},h)$ where $h$ is the induced metric on $\mathcal{S}$.
For the tangent vector $\tilde{v}$ to $\tilde{\gamma}$, it holds that $\tilde{\nabla}_{\tilde{v}}\tilde{v}=0$ along $\tilde{\gamma}$ where $\tilde{v}=v$ at $p$ and $\tilde{\nabla}$ is the covariant derivative associated with $h$.
The geodesic $\tilde{\gamma}$ of $(\mathcal{S},h)$, as the curve of $(M,g)$, has the acceleration $\nabla_{\tilde{v}}\tilde{v}$,
\begin{eqnarray}
\tilde{v}^b\nabla_b\tilde{v}^a
&=&\tilde{v}^b\tilde{\nabla}_b\tilde{v}^a-\epsilon\chi_{bc}\tilde{v}^b\tilde{v}^cn^a\nonumber\\
&=&-\epsilon\chi_{bc}\tilde{v}^b\tilde{v}^cn^a\nonumber\\
&=&-\epsilon\chi_{bc}v^bv^cn^a
\end{eqnarray}
at $p$.
Here, $n^a$ is the unit normal vector of $\mathcal{S}$ and $\epsilon:=n^2$.
This can be interpreted as the acceleration of $\tilde{\gamma}$ with respect to $\gamma$ at $p$.
Therefore, conversely, the geodesic $\gamma$ of $(M,g)$ has the relative acceleration,
\begin{equation}
\llaabel{eq:acceleration-surface}
a_\mathcal{S}^a:=\epsilon\chi_{bc}v^bv^cn^a,
\end{equation}
with respect to $\tilde{\gamma}$, or the hypersurface $\mathcal{S}$, at $p$.
\par
The relative acceleration takes the form,
\begin{equation}
\llaabel{eq:acceleration-surface-null}
a_\mathcal{S}^a=\sigma_{bc}k^bk^cn^a,
\end{equation}
for null vectors $k$ in the case $\mathcal{S}$ is timelike.
From the viewpoint of the relative acceleration $a_\mathcal{S}^a$, a photon surface is a hypersurface on which every (temporally) tangent null geodesics has no relative accelerations with respect to the surface due to the vanishing of $\sigma_{ab}$ at all the point;
$a_\mathcal{S}^a=0$ $\forall\; null \;k\in T_p\mathcal{S}$ $\forall p\in \mathcal{S}$.
\subsection{Reinterpretation of the stability}
The stability of null geodesics on a photon surface, defined in Definition~\ref{definition:stability-gamma}, can be reinterpreted in terms of the relative acceleration $a_S^a$ with the Gaussian normal foliation.
That is, for a photon surface $S$ and the Gaussian normal foliation $\left\{S_r\right\}$, the relative acceleration $a_S^a$ of a perturbed null geodesic $\tilde{\gamma}$ with respect to a nearby hypersurface $\tilde{S}\in \left\{S_r\right\}$ determines whether $\tilde{\gamma}$ is attracted to or repelled from $S$.
\par
Consider a null geodesic $\gamma$ with its tangent vector $k\in T_pS$ at a point $p\in S$.
Let $S_{\delta r}\in\left\{S_r\right\}$ be a hypersurface close to $S$ with a small parameter $\delta r$ and $q\in S_{\delta r}$ be the intersection of $S_{\delta r}$ and the geodesic generated by $n^a$ from $p$.
We generate $k$ from $p$ to $q$ by parallel transport along $n^a$, $\nabla_nk^a=0$.
Then we obtain the nearby null geodesic $\tilde{\gamma}$ with the initial condition $\dot{\tilde{\gamma}}(0)=k\in T_qS_{\delta r}$ at $q$.
The fact $k\in T_qS_{\delta r}$ is guaranteed by the conditions, $n_ak^a=0$ at $p$ and Eq.~(\ref{eq:geodesic-normal-vector}).
If $\gamma$ is stable, i.e. $\tilde{\gamma}$ is attracted to $S$, $\tilde{\gamma}$ has the relative acceleration, $a_{S_{\delta r}}^a$, with respect to $S_{\delta r}$ which is directed toward $S$.
The condition is
\begin{eqnarray}
&&a_{S_{\delta r}}<0\;\;\; for\;\;\;\delta r>0,\nonumber\\
&&a_{S_{\delta r}}>0\;\;\; for\;\;\;\delta r<0
\end{eqnarray}
where $a_{S_{\delta r}}:=n_aa_{S_{\delta r}}^a$.
Taking the limit $\delta r\to 0$, i.e. the limit where $S_r$ and $\tilde{\gamma}$ are infinitesimally close to $S$, the two inequalities reduce to the condition,
\begin{equation}
\left.\nabla_na_{S_{\delta r}}\right|_{\delta r=0}<0.
\end{equation}
In the same way, if $\gamma$ is unstable, the condition is
\begin{equation}
\left.\nabla_na_{S_{\delta r}}\right|_{\delta r=0}>0.
\end{equation}
From Eq.~(\ref{eq:acceleration-surface-null}), the left-hand sides of the conditions further reduce to
\begin{eqnarray}
\llaabel{eq:to-shear-evolution}
\nabla_na_{S_{\delta r}}
&=&\nabla_n\left(k^bk^c\sigma_{bc}\right)\nonumber\\
&=&k^bk^c\nabla_n\sigma_{bc}
\end{eqnarray}
where the last equality is verified by the fact $\nabla_nk^a=0$.
Therefore the normal derivative of the second fundamental form, $\nabla_n\sigma_{ab}$, determines the stability of the photon surface and we indeed reproduce Proposition~\ref{proposition:stability-shear}.
\par

\section{Corollaries}
\llaabel{sec:corollaries}
From the propositions in the previous sections, we can identify the stability of null geodesics on photon surfaces or photon surfaces themselves without specifying the photon surfaces explicitly if the spacetime satisfies some geometrical conditions.
\par
From Proposition~\ref{proposition:stability-curvature}:
\begin{corollary}
\llaabel{corollary:const-curvature}
A photon surface of spacetime of constant curvature is marginally stable.
Specifically, this applies for Minkowski spacetime, de Sitter spacetime, and anti-de Sitter spacetime.

\end{corollary}
\begin{corollary}
The symmetry of spacetime $(M,g)$ and a photon surface $S$ restricts the variation of stability for null geodesics on $S$.
For example, if $S$ is spatially maximally symmetric and also symmetric in a time direction, then all the null geodesics on $S$ has the same stability.
Therefore $S$ is stable, unstable or marginally stable.
Photon spheres are in this case.
\end{corollary}
\par
From Proposition~\ref{proposition:stability-weyl}:
\begin{corollary}
Let $(M,g)$ be a conformally flat spacetime satisfying the null energy condition.
Then a photon surface of $(M,g)$ is stable.
For example, photon surfaces of FLRW spacetime with matter satisfying null energy condition must be stable.
\end{corollary}
\begin{corollary}
Let $(M,g)$ be a spacetime with $\dim M=3$ satisfying the null energy condition.
Then a photon surface of $(M,g)$ is stable.
Therefore, unstable null geodesics are allowed to exist only in spacetime with $\dim M\ge4$ if the null energy condition is satisfied.
For example, photon surfaces of charged rotating BTZ spacetime must be stable.
\end{corollary}
Charged rotating BTZ spacetime is the electrovacuum solution of Einstein-Maxwell equation~\cite{banados_1992}.
If uncharged, BTZ spacetime has constant curvature and Corollary~\ref{corollary:const-curvature} applies.
\begin{corollary}
Let $(M,g)$ and $S$ be a spacetime satisfying the null energy condition and a photon surface of $S$.
Then a null geodesic $\gamma$ in a principal null direction is stable or marginally stable.
\end{corollary}
This is because the principal null condition, $k^bk^ck_{[e}C_{a]bc[d}k_{f]}=0$, implies that only the components corresponding to bases of the form $k^*\otimes \omega$ or $\omega\otimes k^*$ of the second-rank tensor $k^bk^cC_{abcd}$ can be nonzero, where $\omega$ is some one-form and $k^*:=g(\cdot ,k)$ is the one-form dual to the vector $k$~\cite{textbook:wald}. Therefore, the first term in Eq.~(\ref{eq:stabilitycondition-weyl}) vanishes from the fact $n^ak_a=0$ if $k^a$ is in a principal null direction.

\section{Conclusion}
\llaabel{sec:conclusion}
We defined the stability of null geodesics on a photon surface by reformulating the stability of a photon sphere in a covariant manner.
The stability represents whether a null geodesic $\tilde{\gamma}$ perturbed from a null geodesic $\gamma$ on a photon surface is attracted to or repelled from the surface.
Since such a behavior is subject to the geodesic deviation equation, the stability condition of null geodesics on a photon surface is given in terms of Riemann curvature, as in Proposition~\ref{proposition:stability-curvature}, or Weyl and Ricci curvature, as in Proposition~\ref{proposition:stability-weyl}.
We named a photon surface on which all the null geodesics are (un)stable {\it a (un)stable photon surface}.
If there exist no marginally stable null geodesics, the surface is called {\it strictly (un)stable photon surface}.
As we defined the stability only in terms of a local geometrical quantity, the definition is applicable to any photon surfaces even if the photon surfaces and the spacetime have no symmetries.
\par
Although the stability of null geodesics is interpreted as what represents the behavior of perturbed orbits, Proposition~\ref{proposition:stability-curvature} implies that it depends only on the values of the curvature on the photon surface.
This fact can also be seen from our definition, Definition~\ref{definition:stability-gamma}, which requires only the null geodesic and its deviation vector defined just on the surface.
\par
Proposition~\ref{proposition:stability-weyl} tells us that we can a priori identify the stability before finding photon surfaces of spacetime explicitly.
For example, any photon surface in conformally flat spacetime is stable if the null energy condition is satisfied.
Several corollaries concerning this fact were shown in Sec.~\ref{sec:corollaries}.
\par
We also found that the stability of null geodesics can be expressed by the first derivative of the second fundamental form of the surface under an appropriate spacetime foliation, named Gaussian normal foliation.
Proposition~\ref{proposition:stability-shear} might be useful for identifying the stability of a given photon surface.
We demonstrate the calculation of the stability in Appendix~\ref{app:example} for spacetime of spherical, planar and hyperbolic symmetry.

\begin{acknowledgments}
The authors thank J. M. M. Senovilla, D. Ida, C. Yoo, T. Houri, M. Kimura, S. Kinoshita and T. Katagiri for their very helpful discussions and comments.
This work was partially supported by JSPS KAKENHI Grant Number JP19J12007 (Y.K.), JP19K03876 (T.H.).
\end{acknowledgments}

\appendix
\section{Example}
\llaabel{app:example}
We demonstrate the calculation of the stability of a photon surface using Proposition~\ref{proposition:stability-shear}.
Consider the $D$-dimensional spacetime $(M,g)$ with the metric
\begin{equation}
\llaabel{eq:metric-maxsym}
ds^2=-f(r)dt^2+g(r)dr^2+r^2\left(d\chi^2+s^2(\chi)d\Omega^2_{D-3}\right)
\end{equation}
where $f>0$, $g>0$ and
\begin{equation}
d\Omega_{D-3}^2=d\theta_1^2+\cdots+\sin^2\theta_1\cdots\sin^2\theta_{D-5}d\theta_{D-4}^2
+\sin^2\theta_1\cdots\sin^2\theta_{D-4}d\theta_{D-3}^2
\end{equation}
is a unit $(D-3)$-sphere.
The spacetime is static and spherically, planarly or hyperbolically symmetric depending on the function
\begin{equation}
\llaabel{eq:maxsym2space}
s(\chi)=
	\left\{
	\begin{array}{ll}
	\sin \chi& (spherical)\\
	\chi& (planar)\\
	\sinh \chi& (hyperbolic).
	\end{array}
	\right.
\end{equation}
We here investigate a photon surface of constant radius, $r$, which is named {\it constant-$r$ photon surfaces} in~\cite{koga3}.
(The photon surfaces are also called $SO(D-1)\times\mathbb{R}$, $E(D-2)\times\mathbb{R}$ or $SO(1,D-2)\times\mathbb{R}$-invariant photon surfaces depending on the symmetries according to~\cite{claudel}.)
\par
Let $S_r$ be a timelike hypersurface of constant radius, $S_r:=\left\{\left.p\in M\right|r=const.\right\}$, and $n=\sqrt{g}dr$ be the unit normal.
The trace-free part of the second fundamental form is given by
\begin{equation}
\sigma_{(\mu)(\nu)}=-\frac{1}{2(D-1)}\frac{(fr^{-2})'}{(fr^{-2})}\sqrt{g}^{-1}M_{(\mu)(\nu)},\;\;\; M_{(\mu)(\nu)}:=diag\left[D-2,0,1,...,1\right]
\end{equation}
in the tetrad system $\left\{e_{(\mu)}\right\}$ defined so that $e_{(\mu)}\propto\partial_\mu$.
The necessary and sufficient condition for $S_r$ to be a photon surface, $\sigma_{ab}=0$, is equivalent to the condition,
\begin{equation}
\llaabel{eq:const-r-psf}
\left(fr^{-2}\right)'=0.
\end{equation}
We denote the photon surface $S_{r_c}$.
\par
For the constant-$r$ photon surface, the foliation $\left\{S_r\right\}$ in Proposition~\ref{proposition:stability-shear} is a foliation by hypersurfaces of constant radius and here we identify $r$ in $\left\{S_r\right\}$ with $r$ in Eq.~(\ref{eq:metric-maxsym}).
The tensor $\nabla_n{\sigma}_{ab}$ in Proposition~\ref{proposition:stability-shear} is then given by
\begin{eqnarray}
\nabla_n{\sigma}_{(\mu)(\nu)}
&=&\nabla_n{\sigma}_{ab}e_{(\mu)}^ae_{(\nu)}^b \nonumber\\
&=&\nabla_n\left[\sigma_{ab}e_{(\mu)}^ae_{(\nu)}^b\right] -\sigma_{ab}\nabla_n\left[e_{(\mu)}^ae_{(\nu)}^b\right] \nonumber\\
&=&\nabla_n\left[\sigma_{(\mu)(\nu)}\right] \nonumber\\
&=&n^r\partial_r\left[-\frac{1}{2(D-1)}\frac{(fr^{-2})'}{(fr^{-2})}\sqrt{g}^{-1}M_{(\mu)(\nu)}\right] \nonumber\\
&=&-\frac{1}{2(D-1)}\frac{(fr^{-2})''}{(fr^{-2})}g^{-1}M_{(\mu)(\nu)}
\end{eqnarray}
where $n$ is the unit normal vector of $S_{r_c}$ and we used the conditions $\sigma_{ab}=0$ and $\left(fr^{-2}\right)'=0$ on $S_{r_c}$ in the third and last equality, respectively.
Since $k^{(\mu)}k^{(\nu)}M_{(\mu)(\nu)}$ is positive for any null vector $k\in T_pS_{r_c}$, the sign of $k^ak^b\nabla_n{\sigma}_{ab}$ is determined by the factor $(fr^{-2})''$.
Then, from Proposition~\ref{proposition:stability-shear}, the photon surface $S_{r_c}$ is stable, unstable, and marginally stable if and only if
\begin{equation}
\llaabel{eq:const-r-stability}
(fr^{-2})''>0,\ \ <0,\ \ \ and\ =0
\end{equation}
at $r=r_c$, respectively.
\par
The stability condition agrees with that of \cite{koga3}, in which the stability is defined by the effective potential of a null geodesic.

%
\bibliography{stability_psf}

\begin{thebibliography}{16}%
\makeatletter
\providecommand \@ifxundefined [1]{%
 \@ifx{#1\undefined}
}%
\providecommand \@ifnum [1]{%
 \ifnum #1\expandafter \@firstoftwo
 \else \expandafter \@secondoftwo
 \fi
}%
\providecommand \@ifx [1]{%
 \ifx #1\expandafter \@firstoftwo
 \else \expandafter \@secondoftwo
 \fi
}%
\providecommand \natexlab [1]{#1}%
\providecommand \enquote  [1]{``#1''}%
\providecommand \bibnamefont  [1]{#1}%
\providecommand \bibfnamefont [1]{#1}%
\providecommand \citenamefont [1]{#1}%
\providecommand \href@noop [0]{\@secondoftwo}%
\providecommand \href [0]{\begingroup \@sanitize@url \@href}%
\providecommand \@href[1]{\@@startlink{#1}\@@href}%
\providecommand \@@href[1]{\endgroup#1\@@endlink}%
\providecommand \@sanitize@url [0]{\catcode `\\12\catcode `\$12\catcode
  `\&12\catcode `\#12\catcode `\^12\catcode `\_12\catcode `\%12\relax}%
\providecommand \@@startlink[1]{}%
\providecommand \@@endlink[0]{}%
\providecommand \url  [0]{\begingroup\@sanitize@url \@url }%
\providecommand \@url [1]{\endgroup\@href {#1}{\urlprefix }}%
\providecommand \urlprefix  [0]{URL }%
\providecommand \Eprint [0]{\href }%
\providecommand \doibase [0]{http://dx.doi.org/}%
\providecommand \selectlanguage [0]{\@gobble}%
\providecommand \bibinfo  [0]{\@secondoftwo}%
\providecommand \bibfield  [0]{\@secondoftwo}%
\providecommand \translation [1]{[#1]}%
\providecommand \BibitemOpen [0]{}%
\providecommand \bibitemStop [0]{}%
\providecommand \bibitemNoStop [0]{.\EOS\space}%
\providecommand \EOS [0]{\spacefactor3000\relax}%
\providecommand \BibitemShut  [1]{\csname bibitem#1\endcsname}%
\let\auto@bib@innerbib\@empty
\bibitem [{\citenamefont {Synge}(1966)}]{synge}%
  \BibitemOpen
  \bibfield  {author} {\bibinfo {author} {\bibfnamefont {J.~L.}\ \bibnamefont
  {Synge}},\ }\href@noop {} {\bibfield  {journal} {\bibinfo  {journal} {Monthly
  Notices Roy Astron. Soc.}\ }\textbf {\bibinfo {volume} {131}},\ \bibinfo
  {pages} {463} (\bibinfo {year} {1966})}\BibitemShut {NoStop}%
\bibitem [{\citenamefont {Collaboration}\ \emph {et~al.}(2019)\citenamefont
  {Collaboration} \emph {et~al.}}]{eht}%
  \BibitemOpen
  \bibfield  {author} {\bibinfo {author} {\bibfnamefont {E.~H.~T.}\
  \bibnamefont {Collaboration}} \emph {et~al.},\ }\href@noop {} {\bibfield
  {journal} {\bibinfo  {journal} {Astrophys. J. Lett.}\ }\textbf {\bibinfo
  {volume} {875}},\ \bibinfo {pages} {L1} (\bibinfo {year} {2019})},\ \Eprint
  {http://arxiv.org/abs/1906.11238} {arXiv:1906.11238 [astro-ph.GA]}
  \BibitemShut {NoStop}%
\bibitem [{\citenamefont {Cardoso}\ \emph {et~al.}(2009)\citenamefont
  {Cardoso}, \citenamefont {Miranda}, \citenamefont {Berti}, \citenamefont
  {Witek},\ and\ \citenamefont {Zanchin}}]{cardoso}%
  \BibitemOpen
  \bibfield  {author} {\bibinfo {author} {\bibfnamefont {V.}~\bibnamefont
  {Cardoso}}, \bibinfo {author} {\bibfnamefont {A.~S.}\ \bibnamefont
  {Miranda}}, \bibinfo {author} {\bibfnamefont {E.}~\bibnamefont {Berti}},
  \bibinfo {author} {\bibfnamefont {H.}~\bibnamefont {Witek}}, \ and\ \bibinfo
  {author} {\bibfnamefont {V.~T.}\ \bibnamefont {Zanchin}},\ }\href@noop {}
  {\bibfield  {journal} {\bibinfo  {journal} {Phys. Rev. D}\ }\textbf {\bibinfo
  {volume} {79}},\ \bibinfo {pages} {064016} (\bibinfo {year}
  {2009})}\BibitemShut {NoStop}%
\bibitem [{\citenamefont {Hod}(2009)}]{hod}%
  \BibitemOpen
  \bibfield  {author} {\bibinfo {author} {\bibfnamefont {S.}~\bibnamefont
  {Hod}},\ }\href@noop {} {\bibfield  {journal} {\bibinfo  {journal} {Phys.
  Rev. D}\ }\textbf {\bibinfo {volume} {80}},\ \bibinfo {pages} {064004}
  (\bibinfo {year} {2009})}\BibitemShut {NoStop}%
\bibitem [{\citenamefont {Keir}(2016)}]{keir_2014}%
  \BibitemOpen
  \bibfield  {author} {\bibinfo {author} {\bibfnamefont {J.}~\bibnamefont
  {Keir}},\ }\href@noop {} {\bibfield  {journal} {\bibinfo  {journal}
  {Classical Quantum Gravity}\ }\textbf {\bibinfo {volume} {33}},\ \bibinfo
  {pages} {135009} (\bibinfo {year} {2016})},\ \Eprint
  {http://arxiv.org/abs/1404.7036} {arXiv:1404.7036 [gr-qc]} \BibitemShut
  {NoStop}%
\bibitem [{\citenamefont {Cardoso}\ \emph {et~al.}(2014)\citenamefont
  {Cardoso}, \citenamefont {Crispino}, \citenamefont {Macedo}, \citenamefont
  {Okawa},\ and\ \citenamefont {Pani}}]{cardoso_2014}%
  \BibitemOpen
  \bibfield  {author} {\bibinfo {author} {\bibfnamefont {V.}~\bibnamefont
  {Cardoso}}, \bibinfo {author} {\bibfnamefont {L.~C.~B.}\ \bibnamefont
  {Crispino}}, \bibinfo {author} {\bibfnamefont {C.~F.~B.}\ \bibnamefont
  {Macedo}}, \bibinfo {author} {\bibfnamefont {H.}~\bibnamefont {Okawa}}, \
  and\ \bibinfo {author} {\bibfnamefont {P.}~\bibnamefont {Pani}},\ }\href@noop
  {} {\bibfield  {journal} {\bibinfo  {journal} {Phys. Rev. D}\ }\textbf
  {\bibinfo {volume} {90}},\ \bibinfo {pages} {044069} (\bibinfo {year}
  {2014})},\ \Eprint {http://arxiv.org/abs/1406.5510} {arXiv:1406.5510 [gr-qc]}
  \BibitemShut {NoStop}%
\bibitem [{\citenamefont {Cunha}\ \emph {et~al.}(2017)\citenamefont {Cunha},
  \citenamefont {Berti},\ and\ \citenamefont {Herdeiro}}]{cunha}%
  \BibitemOpen
  \bibfield  {author} {\bibinfo {author} {\bibfnamefont {P.~V.~P.}\
  \bibnamefont {Cunha}}, \bibinfo {author} {\bibfnamefont {E.}~\bibnamefont
  {Berti}}, \ and\ \bibinfo {author} {\bibfnamefont {C.~A.~R.}\ \bibnamefont
  {Herdeiro}},\ }\href@noop {} {\bibfield  {journal} {\bibinfo  {journal}
  {Phys. Rev. Lett.}\ }\textbf {\bibinfo {volume} {119}},\ \bibinfo {pages}
  {251102} (\bibinfo {year} {2017})},\ \Eprint
  {http://arxiv.org/abs/1708.04211} {arXiv:1708.04211 [gr-qc]} \BibitemShut
  {NoStop}%
\bibitem [{\citenamefont {Koga}\ and\ \citenamefont {Harada}(2016)}]{koga}%
  \BibitemOpen
  \bibfield  {author} {\bibinfo {author} {\bibfnamefont {Y.}~\bibnamefont
  {Koga}}\ and\ \bibinfo {author} {\bibfnamefont {T.}~\bibnamefont {Harada}},\
  }\href@noop {} {\bibfield  {journal} {\bibinfo  {journal} {Phys. Rev. D}\
  }\textbf {\bibinfo {volume} {94}},\ \bibinfo {pages} {044053} (\bibinfo
  {year} {2016})},\ \Eprint {http://arxiv.org/abs/1601.07290} {arXiv:1601.07290
  [gr-qc]} \BibitemShut {NoStop}%
\bibitem [{\citenamefont {Koga}\ and\ \citenamefont {Harada}(2018)}]{koga2}%
  \BibitemOpen
  \bibfield  {author} {\bibinfo {author} {\bibfnamefont {Y.}~\bibnamefont
  {Koga}}\ and\ \bibinfo {author} {\bibfnamefont {T.}~\bibnamefont {Harada}},\
  }\href@noop {} {\bibfield  {journal} {\bibinfo  {journal} {Phys. Rev. D}\
  }\textbf {\bibinfo {volume} {98}},\ \bibinfo {pages} {024018} (\bibinfo
  {year} {2018})},\ \Eprint {http://arxiv.org/abs/1803.06486} {arXiv:1803.06486
  [gr-qc]} \BibitemShut {NoStop}%
\bibitem [{\citenamefont {Koga}(2019)}]{koga3}%
  \BibitemOpen
  \bibfield  {author} {\bibinfo {author} {\bibfnamefont {Y.}~\bibnamefont
  {Koga}},\ }\href@noop {} {\bibfield  {journal} {\bibinfo  {journal} {Phys.
  Rev. D}\ }\textbf {\bibinfo {volume} {99}},\ \bibinfo {pages} {064034}
  (\bibinfo {year} {2019})},\ \Eprint {http://arxiv.org/abs/1901.02592}
  {arXiv:1901.02592 [gr-qc]} \BibitemShut {NoStop}%
\bibitem [{\citenamefont {Claudel}\ \emph {et~al.}(2001)\citenamefont
  {Claudel}, \citenamefont {Virbhadra},\ and\ \citenamefont {Ellis}}]{claudel}%
  \BibitemOpen
  \bibfield  {author} {\bibinfo {author} {\bibfnamefont {C.~M.}\ \bibnamefont
  {Claudel}}, \bibinfo {author} {\bibfnamefont {K.~S.}\ \bibnamefont
  {Virbhadra}}, \ and\ \bibinfo {author} {\bibfnamefont {G.~F.~R.}\
  \bibnamefont {Ellis}},\ }\href@noop {} {\bibfield  {journal} {\bibinfo
  {journal} {J. Math. Phys.}\ }\textbf {\bibinfo {volume} {42}},\ \bibinfo
  {pages} {818} (\bibinfo {year} {2001})}\BibitemShut {NoStop}%
\bibitem [{\citenamefont {Perlick}(2005)}]{perlick}%
  \BibitemOpen
  \bibfield  {author} {\bibinfo {author} {\bibfnamefont {V.}~\bibnamefont
  {Perlick}},\ }\href@noop {} {\bibfield  {journal} {\bibinfo  {journal}
  {Nonlinear Analysis}\ }\textbf {\bibinfo {volume} {63/5-7}},\ \bibinfo
  {pages} {e511} (\bibinfo {year} {2005})}\BibitemShut {NoStop}%
\bibitem [{\citenamefont {Gibbons}\ and\ \citenamefont
  {Warnick}(2016)}]{gibbons_2016}%
  \BibitemOpen
  \bibfield  {author} {\bibinfo {author} {\bibfnamefont {G.~W.}\ \bibnamefont
  {Gibbons}}\ and\ \bibinfo {author} {\bibfnamefont {C.~M.}\ \bibnamefont
  {Warnick}},\ }\href@noop {} {\bibfield  {journal} {\bibinfo  {journal} {Phys.
  Lett. B}\ }\textbf {\bibinfo {volume} {763}},\ \bibinfo {pages} {169}
  (\bibinfo {year} {2016})}\BibitemShut {NoStop}%
\bibitem [{\citenamefont {Poisson}(2004)}]{textbook:poisson}%
  \BibitemOpen
  \bibfield  {author} {\bibinfo {author} {\bibfnamefont {E.}~\bibnamefont
  {Poisson}},\ }\href@noop {} {\emph {\bibinfo {title} {A Relativist's
  Toolkit}}}\ (\bibinfo  {publisher} {Cambridge University Press, Cambridge,
  United Kingdom},\ \bibinfo {year} {2004})\BibitemShut {NoStop}%
\bibitem [{\citenamefont {Ba{\~{n}}ados}\ \emph {et~al.}(1992)\citenamefont
  {Ba{\~{n}}ados}, \citenamefont {Teitelboim},\ and\ \citenamefont
  {Zanelli}}]{banados_1992}%
  \BibitemOpen
  \bibfield  {author} {\bibinfo {author} {\bibfnamefont {M.}~\bibnamefont
  {Ba{\~{n}}ados}}, \bibinfo {author} {\bibfnamefont {C.}~\bibnamefont
  {Teitelboim}}, \ and\ \bibinfo {author} {\bibfnamefont {J.}~\bibnamefont
  {Zanelli}},\ }\href@noop {} {\bibfield  {journal} {\bibinfo  {journal} {Phys.
  Rev. Lett.}\ }\textbf {\bibinfo {volume} {69}},\ \bibinfo {pages} {1840}
  (\bibinfo {year} {1992})},\ \Eprint {http://arxiv.org/abs/1009.3749}
  {arXiv:1009.3749 [gr-qc]} \BibitemShut {NoStop}%
\bibitem [{\citenamefont {Wald}(1984)}]{textbook:wald}%
  \BibitemOpen
  \bibfield  {author} {\bibinfo {author} {\bibfnamefont {R.~M.}\ \bibnamefont
  {Wald}},\ }\href@noop {} {\emph {\bibinfo {title} {General Relativity}}}\
  (\bibinfo  {publisher} {The University of Chicago Press, Chicago},\ \bibinfo
  {year} {1984})\BibitemShut {NoStop}%
\end{thebibliography}%

\end{document}